\PassOptionsToPackage{unicode}{hyperref}
\PassOptionsToPackage{hyphens}{url}
\PassOptionsToPackage{dvipsnames,svgnames,x11names}{xcolor}
\documentclass[
  a4paper,
]{article}
\usepackage{xcolor}
\usepackage{amsmath,amssymb}
\usepackage{iftex}
\ifPDFTeX
  \usepackage[T1]{fontenc}
  \usepackage[utf8]{inputenc}
  \usepackage{textcomp} 
\else 
  \usepackage{unicode-math} 
  \defaultfontfeatures{Scale=MatchLowercase}
  \defaultfontfeatures[\rmfamily]{Ligatures=TeX,Scale=1}
\fi
\IfFileExists{upquote.sty}{\usepackage{upquote}}{}
\IfFileExists{microtype.sty}{
  \usepackage[]{microtype}
  \UseMicrotypeSet[protrusion]{basicmath} 
}{}
\makeatletter
\@ifundefined{KOMAClassName}{
  \IfFileExists{parskip.sty}{%
    \usepackage{parskip}
  }{
    \setlength{\parindent}{0pt}
    \setlength{\parskip}{6pt plus 2pt minus 1pt}}
}{
  \KOMAoptions{parskip=half}}
\makeatother
\makeatletter
\ifx\paragraph\undefined\else
  \let\oldparagraph\paragraph
  \renewcommand{\paragraph}{
    \@ifstar
      \xxxParagraphStar
      \xxxParagraphNoStar
  }
  \newcommand{\xxxParagraphStar}[1]{\oldparagraph*{#1}\mbox{}}
  \newcommand{\xxxParagraphNoStar}[1]{\oldparagraph{#1}\mbox{}}
\fi
\ifx\subparagraph\undefined\else
  \let\oldsubparagraph\subparagraph
  \renewcommand{\subparagraph}{
    \@ifstar
      \xxxSubParagraphStar
      \xxxSubParagraphNoStar
  }
  \newcommand{\xxxSubParagraphStar}[1]{\oldsubparagraph*{#1}\mbox{}}
  \newcommand{\xxxSubParagraphNoStar}[1]{\oldsubparagraph{#1}\mbox{}}
\fi
\makeatother

\usepackage{longtable,booktabs,array}
\usepackage{calc} 
\usepackage{etoolbox}
\makeatletter
\patchcmd\longtable{\par}{\if@noskipsec\mbox{}\fi\par}{}{}
\makeatother
\IfFileExists{footnotehyper.sty}{\usepackage{footnotehyper}}{\usepackage{footnote}}
\makesavenoteenv{longtable}
\usepackage{graphicx}
\makeatletter
\newsavebox\pandoc@box
\newcommand*\pandocbounded[1]{
  \sbox\pandoc@box{#1}%
  \Gscale@div\@tempa{\textheight}{\dimexpr\ht\pandoc@box+\dp\pandoc@box\relax}%
  \Gscale@div\@tempb{\linewidth}{\wd\pandoc@box}%
  \ifdim\@tempb\p@<\@tempa\p@\let\@tempa\@tempb\fi
  \ifdim\@tempa\p@<\p@\scalebox{\@tempa}{\usebox\pandoc@box}%
  \else\usebox{\pandoc@box}%
  \fi%
}
\def\fps@figure{htbp}
\makeatother

\NewDocumentCommand\citeproctext{}{}

\makeatletter
 \let\@cite@ofmt\@firstofone
 \def\@biblabel#1{}
 \def\@cite#1#2{{#1\if@tempswa , #2\fi}}
\makeatother
\newlength{\cslhangindent}
\newlength{\csllabelwidth}
\newenvironment{CSLReferences}[2] 
 {\begin{list}{}{%
  \setlength{\itemindent}{0pt}
  \setlength{\leftmargin}{0pt}
  \setlength{\parsep}{0pt}
  \ifodd #1
   \setlength{\leftmargin}{\cslhangindent}
   \setlength{\itemindent}{-1\cslhangindent}
  \fi
  \setlength{\itemsep}{#2\baselineskip}}}
 {\end{list}}
\usepackage{calc}

\newcommand{\CSLLeftMargin}[1]{\parbox[t]{\csllabelwidth}{\strut#1\strut}}
\newcommand{\CSLRightInline}[1]{\parbox[t]{\linewidth - \csllabelwidth}{\strut#1\strut}}

\usepackage{arxiv}
\usepackage{orcidlink}
\usepackage{amsmath}
\usepackage[T1]{fontenc}
\makeatletter
\@ifpackageloaded{caption}{}{\usepackage{caption}}
\AtBeginDocument{%
\ifdefined\contentsname
  \renewcommand*\contentsname{Table of contents}
\else
  \newcommand\contentsname{Table of contents}
\fi
\ifdefined\listfigurename
  \renewcommand*\listfigurename{List of Figures}
\else
  \newcommand\listfigurename{List of Figures}
\fi
\ifdefined\listtablename
  \renewcommand*\listtablename{List of Tables}
\else
  \newcommand\listtablename{List of Tables}
\fi
\ifdefined\figurename
  \renewcommand*\figurename{Figure}
\else
  \newcommand\figurename{Figure}
\fi
\ifdefined\tablename
  \renewcommand*\tablename{Table}
\else
  \newcommand\tablename{Table}
\fi
}
\@ifpackageloaded{float}{}{\usepackage{float}}
\floatstyle{ruled}
\@ifundefined{c@chapter}{\newfloat{codelisting}{h}{lop}}{\newfloat{codelisting}{h}{lop}[chapter]}
\floatname{codelisting}{Listing}

\makeatother
\makeatletter
\@ifpackageloaded{caption}{}{\usepackage{caption}}
\@ifpackageloaded{subcaption}{}{\usepackage{subcaption}}
\makeatother
\usepackage{bookmark}
\IfFileExists{xurl.sty}{\usepackage{xurl}}{} 
\hypersetup{
  pdftitle={From Prompts to Preferences: An Open-Source Platform for Generative AI-Enhanced Conjoint Analysis},
  pdfauthor={Philipp Brauner},
  pdfkeywords={conjoint analysis, discrete choice experiment, generative
AI, science and technology studies, marketing research, stimuli
generation, survey methodology},
  colorlinks=true,
  linkcolor={blue},
  filecolor={Maroon},
  citecolor={Blue},
  urlcolor={Blue},
  pdfcreator={LaTeX via pandoc}}

\renewcommand{\today}{2026-06-11}
\newcommand{\runninghead}{A Preprint }
\title{From Prompts to Preferences: An Open-Source Platform for
Generative AI-Enhanced Conjoint Analysis}
\author{\textbf{Philipp
Brauner}~\orcidlink{0000-0003-2837-5181}\\Communication Science\\RWTH
Aachen
University\\\\\href{mailto:brauner@comm.rwth-aachen.de}{brauner@comm.rwth-aachen.de}}
\date{2026-06-11}
\begin{document}
\maketitle
\begin{abstract}
Conjoint analysis is a widely used preference measurement method in
marketing research, political science, healthcare, and human-computer
interaction. Despite broad adoption, researchers without access to
commercial platforms face significant barriers, as existing tools are
either expensive or lack end-to-end survey infrastructure. This paper
presents an open-source, self-hosted web application for designing,
deploying, and analysing conjoint surveys. Beyond conventional tabular
stimuli, the platform uses generative AI to produce integrated stimuli
formats: textual scenario descriptions generated by a large language
model, and visual stimuli by a text-to-image model. A researcher-defined
base prompt is parameterised with the conjoint profile, and optional
LLM-facing level annotations enrich the generation. A structured setup
wizard, AI-assisted attribute suggestion, and live data analysis lower
the technical barriers for researchers new to conjoint methodology. A
full export bundle including all stimuli, their generating prompts, and
response data facilitates transparency and reproducibility. The platform
is demonstrated through a proof-of-concept study on care robot
preferences for ambient assisted living (AAL, N=55) using AI-generated
visual stimuli. The paper discusses the role of AI assistance in
conjoint design, arguing that theoretical grounding must remain the
researcher's responsibility, and outlining how genAI-generated stimuli
can broaden the methodological repertoire for HCI and related fields.
\end{abstract}
{\bfseries \emph Keywords}
\def\sep{\textbullet\ }
conjoint analysis \sep discrete choice experiment \sep generative
AI \sep science and technology studies \sep marketing
research \sep stimuli generation \sep 
survey methodology

\section{Introduction}\label{introduction}

Understanding how people form preferences and make trade-offs is a
central concern across disciplines from marketing and economics to
political science {[}1{]}, public health {[}2{]}, human-computer
interaction {[}3{]}, and decarbonization research {[}4{]}. Conjoint
analysis, a family of stated-preference methods in which respondents
choose repeatedly between systematically varied alternatives, has become
one of the most widely used quantitative tools for eliciting such
preferences {[}5{]}, {[}6{]}. The method's appeal lies in its approach:
by orthogonally varying the attributes of competing options, conjoint
designs allow researchers to efficiently decompose overall preferences
into so-called ``part-worth utilities'' associated with individual
attribute levels, recovering the implicit weights people place on
different features when making trade-off decisions.

Despite its methodological maturity, access to conjoint survey
infrastructure remains unevenly distributed. Commercial platforms such
as Sawtooth Software's Lighthouse Studio {[}7{]} are the de-facto
standard for professional and applied conjoint research, but their
licensing costs place them beyond the reach of many academic research
groups, particularly in disciplines, such as human-computer interaction,
where conjoint is not a standard method and departmental budgets are not
organised around it. General-purpose survey tools such as Qualtrics or
SoSci Survey offer limited conjoint support, but these either require
specialised expertise to configure or do not support the full range of
design types. Statistical packages for R, including \texttt{idefix}
{[}8{]} and \texttt{support.CEs} {[}9{]}, provide rigorous design
generation and analysis capabilities but do not supply survey delivery
infrastructure: researchers must separately implement participant
routing, response collection, and data management. No freely available,
end-to-end, self-hosted platform currently covers the lifecycle from
survey design through participant data collection to export of
analysis-ready data.

Consequently, conjoint analysis---and the precise preference measurement
it affords---is in practice available primarily to researchers at
well-funded institutions or to those with substantial programming
skills. This prevents broader methodological adoption by smaller
research groups and in emerging areas such as the study of public
preferences, interface features, technology design, or privacy
trade-offs in interactive systems.

Beyond the access problem, most conventional conjoint studies rely on
tabular stimuli: alternatives are displayed as a table in which rows
correspond to attributes and columns to alternatives, and participants
inspect the cells to form their preferences. This format prioritises
comparability and attribute salience, but it may not be well-suited to
all research contexts. Preferences for complex products, services, or
social situations may be better elicited through richer, integrated
stimuli that allow participants to engage with alternatives as coherent
wholes rather than as bundles of abstract attribute-level labels. Even
when integrated stimuli could be produced, doing so was cumbersome.
Recent advances in large language models (LLMs) and text-to-image
generation now make it technically and economically feasible to generate
such stimuli at scale from structured profiles.

This paper makes three contributions. First, we present
GenerativeConjoint, an open-source conjoint survey platform that covers
research design and stimuli generation, survey delivery, and export of
analysis-ready data files. Second, we extend the conventional tabular
format with two generative AI stimuli modes: a textual mode in which
LLM-generated scenario descriptions instantiate each conjoint profile,
and a visual mode in which text-to-image generation produces synthetic
images representing each alternative. Third, we reflect on the
(in)appropriate uses of generative AI assistance in conjoint research
workflows, arguing that AI assistance for stimuli generation holds
genuine promise, while AI-generated attribute and level structures must
always be grounded in domain theory and validated by the researcher.

\section{Background}\label{background}

\subsection{Conjoint Analysis and Discrete Choice
Experiments}\label{conjoint-analysis-and-discrete-choice-experiments}

Conjoint analysis originated in mathematical psychology and was advanced
and popularized in marketing research by Green and Srinivasan {[}5{]}.
The method derives from the idea that complex stimuli can be understood
as combinations of attributes, and that overall evaluations of such
stimuli can be decomposed into the separate contributions of each
attribute level. Early implementations relied on full-profile ratings,
but the field progressively shifted toward choice-based conjoint (CBC),
in which respondents choose their preferred alternative from a set of
simultaneously presented profiles {[}10{]}. CBC, also called discrete
choice experiment (DCE), has become the standard paradigm: it mirrors
purchasing or decision-making behaviour more closely than rating tasks,
yields data directly analysable under random utility theory, and
supports mixed logit and latent class models that account for preference
heterogeneity across respondents {[}11{]}.

A conjoint design is defined by a set of attributes (the dimensions
along which alternatives vary), each with a discrete set of levels (the
values an attribute can take). A full factorial design crossing all
attributes and levels would grow combinatorially and is impractical to
administer. Instead, experimental design theory is used to select a
subset of profiles (typically using D-optimal or orthogonal designs)
that permit efficient estimation of part-worth utilities {[}12{]}. The
orthogonal or near-orthogonal structure ensures that attribute effects
can be estimated independently of one another. Respondents answer a set
of choice tasks, each presenting two or more alternatives {[}6{]}.

\subsection{Applications of Conjoint
Analysis}\label{applications-of-conjoint-analysis}

Conjoint analysis has been applied productively across a wide range of
disciplines. In marketing and consumer research, conjoint studies have
been used to estimate willingness to pay, optimise product features, and
segment markets {[}13{]}. In health economics and medical
decision-making, discrete choice experiments elicit patient and
clinician preferences for treatment options, service configurations, and
health outcomes {[}14{]}, {[}15{]}. In political science, conjoint
survey experiments have become a central tool for studying complex
attitude formation, most prominently in research on immigration
preferences and candidate evaluation {[}16{]}, {[}17{]}. The analytical
framework of average marginal component effects (AMCEs) developed by
Hainmueller, Hopkins, and Yamamoto {[}16{]} has substantially influenced
how political scientists design and analyse conjoint studies.

In human-computer interaction and science and technology studies,
conjoint analysis has been used to study tradeoffs and the relatative
weightings of features across a broad range of domains. Examples include
privacy and trust trade-offs in smart home technology {[}18{]}, moral
consideration of artificial intelligences based on features such as
embodiment, autonomy, and emotion expression {[}19{]}, features of care
robots for older adults {[}20{]}, trade-offs in workplace monitoring
technologies {[}21{]}, and user interface preferences for generative AI
systems {[}22{]}. Given that HCI research increasingly grapples with how
users evaluate AI-assisted tools and generative systems, weighing
accuracy against transparency, convenience against privacy, autonomy
against control, conjoint analysis may offer an approach to quantifying
these preference structures. Yet precisely because conjoint is not yet a
default method in many HCI research groups, the barrier to adoption
matters.

\subsection{Stimuli Presentation in Conjoint
Studies}\label{stimuli-presentation-in-conjoint-studies}

Conventional conjoint studies often present profiles in tabular form.
Each alternative is a column, each attribute is a row, and the cell
content is the level label. This format is transparent and easy to
compare, but it has known limitations. It may prime respondents to
process attributes independently, suppressing holistic or gestalt-level
evaluation that would arise in natural decision contexts {[}23{]}. It
treats all attributes as commensurable, which may not hold when
alternatives represent complex sociotechnical scenarios. And it can lead
to satisficing or attribute non-attendance, in which respondents focus
on one or two salient attributes and ignore others {[}24{]}.

Alternative stimuli formats have been explored in the literature.
Pictorial or image-based conjoint studies have been used in marketing,
such as product design and food preference research, showing that richer
visual stimuli may reduce cognitive load and increase ecological
validity {[}23{]}, {[}25{]}. Scenario-based vignette studies present
alternatives as short textual descriptions, a format well-established in
sociology and political science {[}26{]}. However, creating textual or
visual stimuli manually for all shown profiles is laborious and
expensive, which has historically limited these formats to small-scale
studies. Generative AI now makes it feasible to produce such stimuli
programmatically from structured profile data, opening new
methodological possibilities.

\section{Generative AI supported Conjoint
Platform}\label{generative-ai-supported-conjoint-platform}

The tool described in the following (hereafter referred to as
GenerativeConjoint) was designed around several objectives.

\emph{Accessibility: } The platform should be usable without specialised
statistical or programming knowledge. Researchers should be able to
configure a full conjoint study through a web interface, without writing
code or managing statistical design software separately.

\emph{Methodological rigour:} The underlying design generation must
produce statistically sound designs: near-orthogonal, approximately
level-balanced, with quantified D-efficiency {[}6{]}, but without
overwhelming the user. Participant routing, session management, and
response storage should be robust to partial completions and re-entry.

\emph{Flexible stimuli formats: } Beyond the conventional tabular
presentation, the platform should support integrated textual and visual
stimuli generated from conjoint profiles via generative AI, enabling
researchers to study preference formation in richer representational
contexts.

\emph{Research data management: } All materials necessary for
reproducible analysis and archival---response data, the design matrix,
stimuli content, as well as the prompts used to generate
stimuli---should be exportable in structured, analysis-ready formats.

\emph{Open source and self-hosted: } The platform must be freely
available, require no subscription or per-response fees, and be
deployable on institutional infrastructure without dependency on
third-party data processors beyond an (optional) API key for genAI
services.

\subsection{Survey Design and
Configuration}\label{survey-design-and-configuration}

The platform supports conjoint surveys with 1--6 attributes and 2--4
levels per attribute. These limits reflect established practical
guidance: beyond six attributes, respondent fatigue increases sharply
and the required sample size grows substantially {[}6{]}. The number of
choice tasks (\(T\)) and alternatives per task (\(J\)) are configurable
within validated ranges and a sample size estimator based on the Orme
heuristic provides real-time guidance as researchers configure their
design {[}6{]}.

Technically, the experimental design is generated using a coordinate
exchange algorithm that maximises the D-efficiency of the information
matrix. D-efficiency measures the precision of part-worth utility
estimates relative to an orthogonal benchmark; a D-efficient design
minimises the volume of the confidence ellipsoid around the estimated
parameters {[}27{]}. All attribute-level combinations are encoded using
effects coding (sum-to-zero constraint), consistent with standard
conjoint analysis practice.

\subsection{Presentation Modes}\label{presentation-modes}

Surveys can have one of three stimuli presentation modes.

\emph{Tabular} presents the standard conjoint format: a table with
attribute rows and alternative columns. Each cell contains the level
label and description for that attribute. As with other software
solutions, the selection at the bottom of the table allows participants
to indicate their preferred design alternative.

\emph{Textual mode} replaces the table with AI-generated scenario
descriptions. The researcher provides a base prompt with a set of
content or stylistic instructions written for the language model. The
system automatically appends the attribute-level profile for each
alternative as a structured list. For each of the profiles, the platform
calls a text generator\footnote{At the time of writing, OpenAI's GPT-4o
  mini.} to generate the scenario text (typically 3--5 sentences) that
embeds the profile's characteristics in a coherent narrative. During
setup, each level can optionally carry an LLM-facing annotation (a
``hint'') that provides additional context to the model without being
shown to participants. For example, a level labelled ``High screen
brightness'' might carry the hint ``1000 nits peak brightness, suitable
for outdoor use in direct sunlight'', giving the language model concrete
detail while keeping the participant-facing label concise. In the
survey, the generated texts are displayed to participants as labelled
cards, one per alternative profile. Advanced users may also modify the
base prompt to generate texts to contain only a subset of the attributes
and using the other attributes to modify the linguistic properties of
the generated texts. Example use cases are the study of the
persuasiveness of different narratives, framings, or communication
styles.

\emph{Visual mode} follows the same parametric generation logic but uses
a text-to-image model\footnote{At the time of writing, OpenAI's
  gpt-image-2 model accessed via the Images API.} to produce synthetic
images for each profile. Again, the researcher provides a base image
prompt describing the scene type, visual style, and composition, and the
profile details, including level hints, are appended as a structured
suffix before the prompt is sent to the image generation API. The survey
tool then presents just the stimuli without any text and participants
select their preferred alternative by clicking on its image.

In both textual and visual modes, AI-assisted prompt optimisation is
available: an ``Optimize for image generation'' function calls a
language model to rewrite the base prompt into a form better suited to
image generation; a feedback function evaluates the combined prompt
against a randomly sampled profile to allow iterative refinement before
committing to full, lengthy, and potentially costly text or image
generation.

\subsection{Export and Research Data
Management}\label{export-and-research-data-management}

Reproducibility and research data management were considered in the
platform's design. An export function bundles all relevant research data
as a single ZIP archive. Beyond the participants' responses (directly
usable for R packages for conditional logit (\texttt{survival::clogit})
and mixed logit (\texttt{mlogit::mlogit}) estimations), the file
contains 1) the pre-generated design matrix with one row per task and
alternative profiles, containing all attribute-level assignments, 2) a
structured Markdown document recording survey metadata, the
attribute-level table with all descriptions and level annotations, the
redirect URL template, and a data dictionary for the response data, 3) a
starter analysis script that reads the response CSV, coerces attribute
columns to ordered factors, and provides commented examples for common
model specifications, 4) a table containing all attribute-level
assignments, the full prompts sent to the language model, and the
generated scenario texts, 5) all prompts for image generation and an
images folder containing the full-resolution PNGs for each profile (see
data repository of the validation study for an example).

For text and image tasks, archiving the prompts is essential for
reproducibility: the same prompt may produce different outputs at
different times or with updated models. Also, the prompt encodes the
researcher's methodological choices about framing, register, and
narrative style. The data export automatically provides a complete audit
trail of how each stimulus was produced. Together, these files allow
complete documentation of the visual stimuli in line with research data
management requirements.

\subsection{The Survey Creator View}\label{the-survey-creator-view}

After login, the survey administrators see a list of available surveys
and the option to setup a new survey. Instead of an self-developed user
management, researchers log in with their existing ORCID ID and surveys
can be shared between users using their ORCIDs. After creating or
opening a project, the users are guided through the setup following the
\emph{wizard} design pattern {[}28{]}.

\emph{Step 1: Basic Information.} The researcher enters the survey
title, language, and a description of the research context. Though the
description is not shown to participants, it may be used for the
AI-based attribute and level suggestion and as input to the automated
participant introduction generator.

\begin{figure}

\begin{minipage}[t]{0.47\linewidth}

\centering{

\pandocbounded{\includegraphics[keepaspectratio]{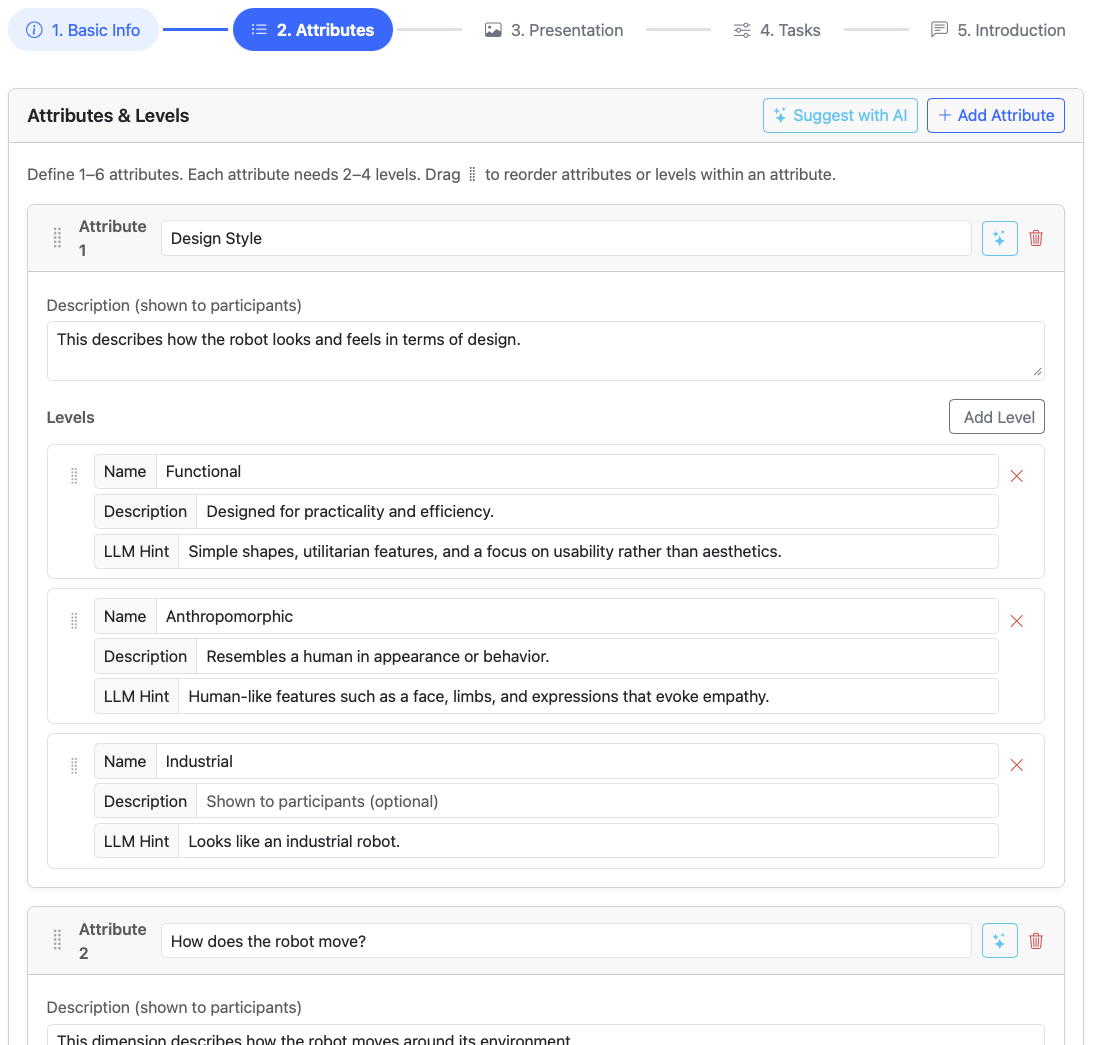}}

}

\subcaption{\label{fig-creator-a}Attribute and level configuration (Step
2), showing the drag-and-drop interface with participant-facing
descriptions and optional LLM-facing hints per level.}

\end{minipage}%
\begin{minipage}[t]{0.06\linewidth}
~\end{minipage}%
\begin{minipage}[t]{0.47\linewidth}

\centering{

\pandocbounded{\includegraphics[keepaspectratio]{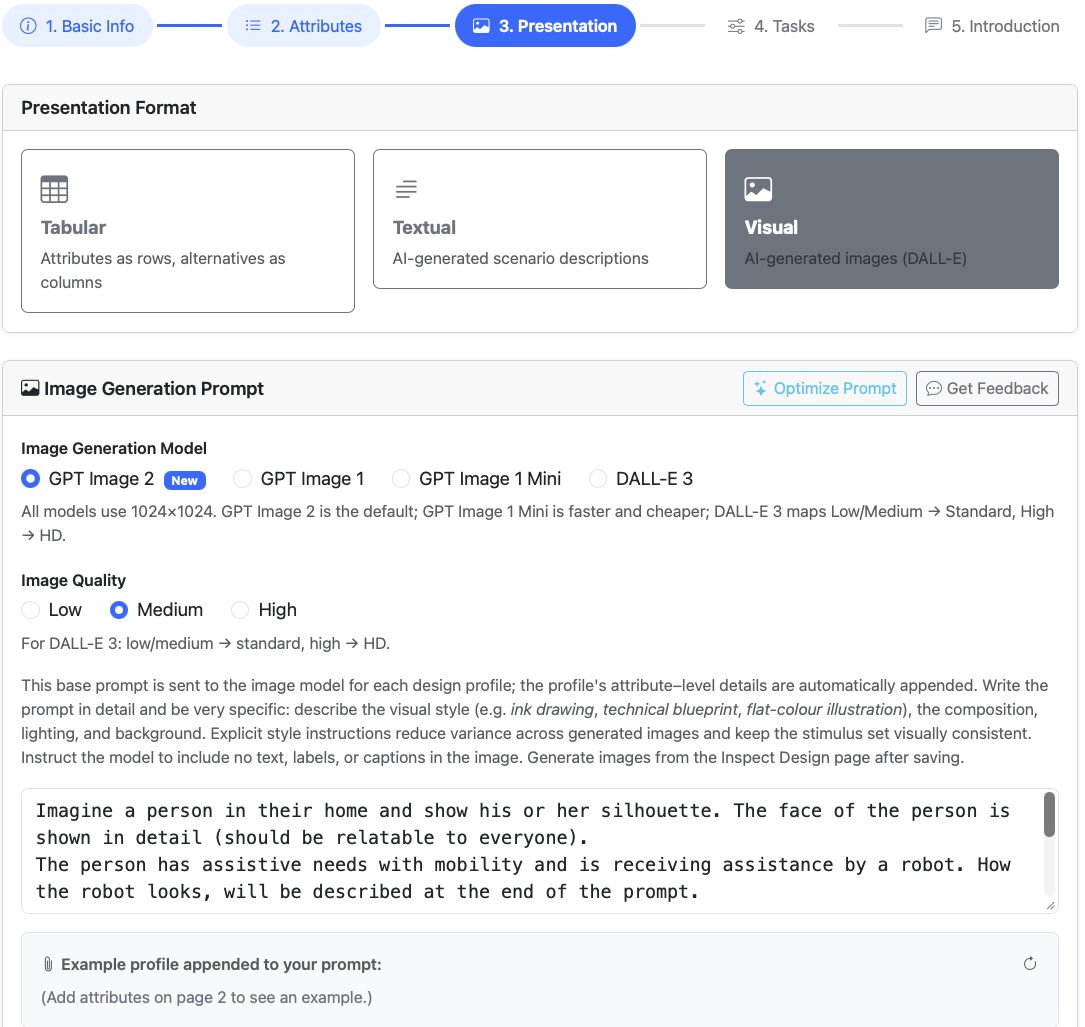}}

}

\subcaption{\label{fig-creator-b}Image generation prompt interface (Step
3), showing the base prompt editor with AI-assisted optimisation and a
preview of a generated stimulus.}

\end{minipage}%

\caption{\label{fig-creator}Screenshots of the GenerativeConjoint survey
configuration wizard.}

\end{figure}

\emph{Step 2: Attributes and Levels.} Attributes are added, named, and
populated with 2--4 levels using a drag-and-drop interface. Each
attribute and each level can carry a participant-facing description
explaining the dimension and design option. Each level can further
contain an LLM-facing annotation used to support text or image stimuli
generation. Figure~\ref{fig-creator-a} shows the interface for
configuring attributes and levels.

Attributes and levels for the study, as well as the generation of the
participant and LLM facing descriptions can be generated via a language
model, substantially reducing the time needed to configure complex
surveys. Obviously, AI-suggested attributes and levels, as well as the
descriptions should not be adopted uncritically, as the structure of a
conjoint study should build on theoretical commitments and must be
grounded in prior theory and domain knowledge (see Discussion).

\emph{Step 3: Presentation and Configuration.} The researcher selects
the presentation mode (tabular, textual, or visual) and configures the
number of tasks and alternatives. For all modes, contextualised
configuration cards appear in which the presentation of the attributes
can be configured or the relevant generation base prompt can be written,
evaluated with LLM support, and refined. Figure~\ref{fig-creator-b}
illustrates the image generation prompt interface.

\emph{Step 4: Participant Introduction.} An optional introductory text
is shown to participants before the first task. A prompt template
pre-filled with the survey's attributes and task configuration can be
sent to a language model to draft the introduction, which the researcher
then reviews and edits. As with attribute suggestions, in real research
settings this text should be defined and checked before the survey is
created.

After saving, the platform generates the conjoint design and displays a
survey detail page. From this page the researcher can inspect the
attribute structure, the generated design, participant statistics
(started and completed), and the entry and exit links for embedding in
an external survey platform (such as \emph{SoSci Survey} or
\emph{LimeSurvey}). The full data export is also available from this
page. A dedicated Inspect Design page offers a tabular view of the full
design matrix organised by block, and a stimuli tab from which text or
image generation can be triggered: either for all missing profiles at
once or for individual profiles, allowing researchers to test prompt
feasibility before committing to full-scale generation. Should
individual stimuli not match the researchers' quality criteria, they can
be deleted and regenerated using the same prompt.

A dedicated results page provides a live (though preliminary) view of
preference estimates and survey statistics. As data accumulate and the
survey runs, mean and median completion time, completed and dropped-out
cases, as well as the relative attribute importance and per-level
part-worth utilities are computed from response data. The results page
is accessible via a shareable token-authenticated URL, allowing
researchers to give collaborators or stakeholders a read-only view of
accumulating results without granting access to the full survey
configuration.

\subsection{What Survey Participants
See}\label{what-survey-participants-see}

Participants arrive at the platform via a URL carrying a unique
participant key, typically redirected from an external survey platform
that handles consent, screening, demographics, and personality measures.
GenerativeConjoint receives the key and presents the survey tasks in
sequence.

If the survey has an (optional) introduction text, participants see it
on a dedicated page before the first task. The task interface is
deliberately minimal: a progress bar indicating the current task out of
the total, a brief instruction prompt, and the choice alternatives.

In tabular mode, the alternatives are displayed as a standard conjoint
table. In textual mode, alternatives are displayed as styled cards, each
containing the generated scenario text. In visual mode, alternatives
image cards are displayed. In all modes, clicking a design option (or a
``select'' indicator) records the choice and a brief visual confirmation
is provided before advancing automatically.

After the final task, participants see a completion page and are
redirected to the external survey platform using the configurable URL
that also carries the participants' unique key if desired. This closes
the loop for the host survey and allows linking conjoint responses to
other survey items collected on the external platform.

\subsection{Technical Realisation}\label{technical-realisation}

The platform is implemented as a Flask 3 application with SQLAlchemy and
a SQLite database. The choice of SQLite as the default database reflects
the self-hosted research context: it requires no separate database
server process and stores the entire survey dataset as a single file
that is straightforward to back up and archive and sufficiently
performant for most research tasks. Of course, the database is
configurable, allowing migration to dedicated databases for faster
performance. The frontend uses Bootstrap 5 for layout and styling,
SortableJS for drag-and-drop reordering of attributes and levels, and
vanilla JavaScript for all dynamic behaviour.

The conjoint design generation is implemented in a pure Python module
using NumPy for matrix operations. The coordinate exchange algorithm
operates directly on the effects-coded design matrix, iterating over all
(task, alternative, attribute) positions and testing all possible level
assignments, accepting any change that increases the determinant of the
information matrix. Multiple random restarts are run in sequence; the
design with the highest D-efficiency is used.

Generative AI-based stimuli generation uses the OpenAI API. Text
generation calls the Chat Completions API with GPT-4o mini, a
cost-efficient configuration appropriate for large-scale batch
generation where immediate turnaround is not required. Image generation
supports different image models via the Images API; researchers can
select different quality levels to balance output fidelity against cost.
Stimuli are stored per unique attribute-level profile: when the same
profile occurs in multiple blocks of a blocked design, a single
generated stimulus is shared across all occurrences, reducing generation
cost and ensuring stimulus consistency across blocks.

\section{Proof of Concept Study: Preferences of Care Robot
Designs}\label{proof-of-concept-study-preferences-of-care-robot-designs}

To demonstrate the platform, a study on the preferences for robot
designs in the context of Ambient Assisted Living was conducted. Care
robots should support people with needs in their everyday lives {[}29{]}
and design aesthetics and interaction style have been identified as
important factors of acceptance in stakeholder preference studies
{[}20{]}, {[}30{]}. The study used GenerativeConjoint's visual
presentation mode with four attributes and three levels each:
\emph{Design Style} (Functional, Anthropomorphic, Industrial),
\emph{Locomotion} (Wheels, Legs, Tracks), \emph{Size} (Small, Medium,
Large), and \emph{Location} (Bedroom, Living Room, Kitchen).

\begin{figure}

\begin{minipage}[t]{0.45\linewidth}

\centering{

\pandocbounded{\includegraphics[keepaspectratio]{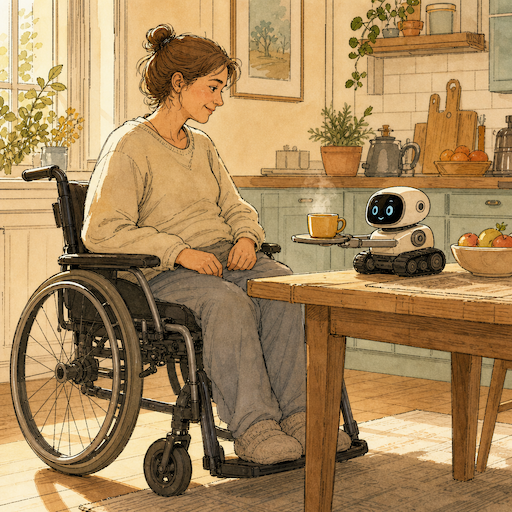}}

}

\subcaption{\label{fig-stim-a}Anthropomorphic design, tracks, small,
kitchen}

\end{minipage}%
\begin{minipage}[t]{0.10\linewidth}
~\end{minipage}%
\begin{minipage}[t]{0.45\linewidth}

\centering{

\pandocbounded{\includegraphics[keepaspectratio]{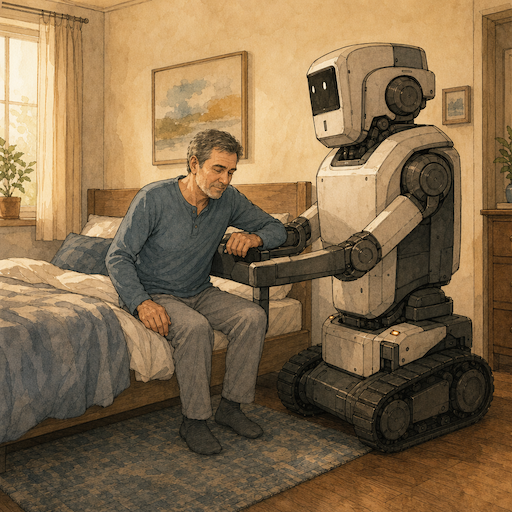}}

}

\subcaption{\label{fig-stim-b}Industrial design, tracks, large, bedroom}

\end{minipage}%
\newline
\begin{minipage}[t]{0.45\linewidth}

\raisebox{-\height}{

\pandocbounded{\includegraphics[keepaspectratio,alt={Industrial design, legs, medium size, kitchen}]{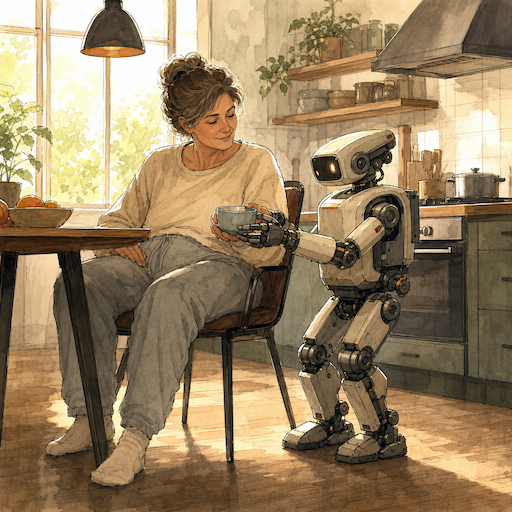}}

}

\subcaption{\label{}Industrial design, legs, medium size, kitchen}
\end{minipage}%
\begin{minipage}[t]{0.10\linewidth}
~\end{minipage}%
\begin{minipage}[t]{0.45\linewidth}

\raisebox{-\height}{

\pandocbounded{\includegraphics[keepaspectratio,alt={Functional design, wheels, large, living room}]{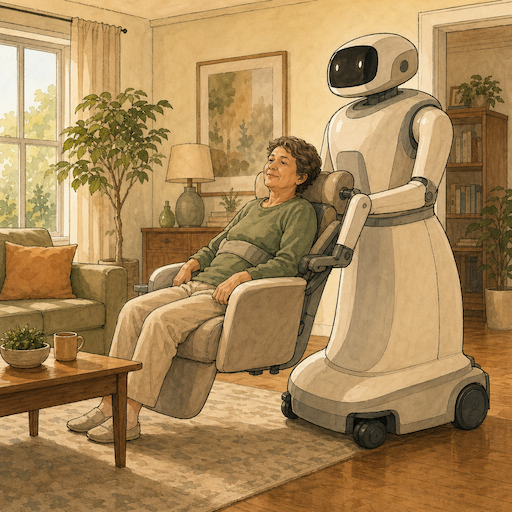}}

}

\subcaption{\label{}Functional design, wheels, large, living room}
\end{minipage}%

\caption{\label{fig-stimuli}AI-generated stimuli (with `gpt-image-2')
from the robots for ambient assisted living study. Each image
instantiates a distinct attribute-level profile; both share the same
ink-drawing base prompt. All stimuli and prompts are available in the
data repository.}

\end{figure}

Each participant completed eight forced-choice tasks, each presenting
two simultaneously displayed AI-generated robot images with different
features. The fractional factorial design (with participants evaluating
only a small share of all possible design pairs) achieved a
\(D\)-efficiency of 0.61. Images were generated with OpenAI's
\texttt{gpt-image-2} model, the most recently released text-to-image
model at the time of the study. The shared base prompt instructed the
model to produce an ink drawing in light, friendly, and warm colours
with a hand-drawn aesthetic; the depicted person was described as
relaxed, with care needs evident in their posture.
Figure~\ref{fig-stimuli} shows two representative stimuli from the
study, illustrating how the attribute-level profile is translated into a
coherent visual scene.

After the conjoint study, the usability and suitability of the software
was evaluated using seven items measured on a 5-point Likert scale using
SoSci Survey (1 = Strongly Disagree, 5 = Strongly Agree). The items
focused on three primary aspects: 1) System Usability: Ease of
navigation and interface utility. 2) Visual Quality: Clarity, aesthetic
appeal, and communicative power of the imagery. 3) Contextual
Integration and scenario utility: The ability to visualize the presented
technology within personal environments and the perceived value of using
images to present the scenarios.

\emph{Description of the sample:} The study was fielded via Prolific,
recruiting 55 participants (25 women, 28 men; age range 19--79 years
with a median age of 44 years) residing in the United Kingdom.
Participation was voluntary and participants received a small financial
compensation of approximately €0.50. The study was conducted in
accordance with the ethical guidelines of the German Psychological
Society (DGPs); formal ethics committee review was not required given
the anonymous, low-risk nature of the online survey and the absence of
medical or clinical interventions. Informed consent was obtained from
all participants prior to the study via the platform's built-in consent
screen.

\begin{figure}

\centering{

\pandocbounded{\includegraphics[keepaspectratio]{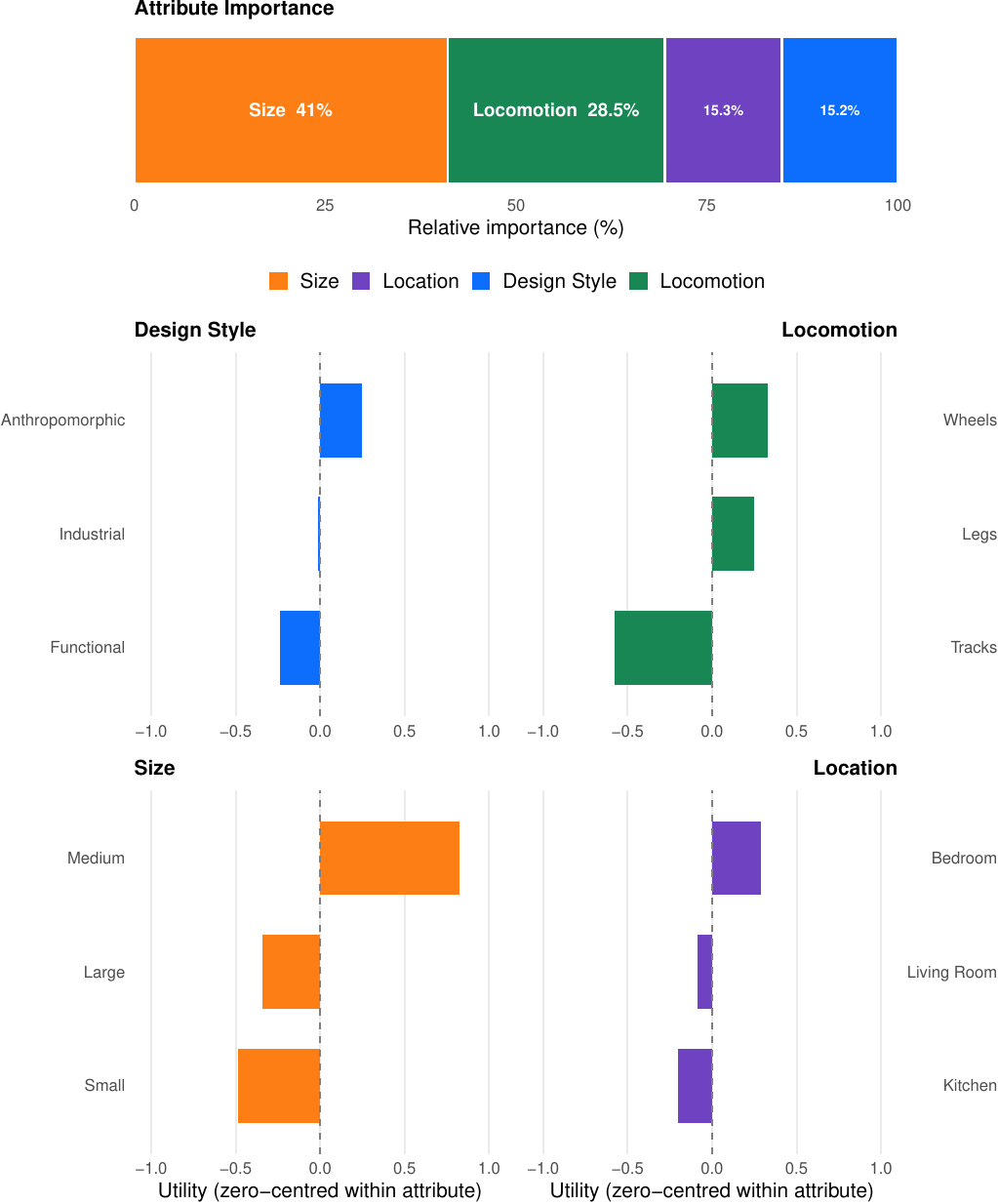}}

}

\caption{\label{fig-results}Attribute importance (top) and part-worth
utilities (bottom) for the robots for ambient assisted living conjoint
study (conditional logit, N = 55, zero-centred within attribute).}

\end{figure}

\emph{Results:} Robot size emerged as the most important attribute
(41.0\%), followed by the mode of locomotion (28.5\%), room location
(15.3\%), and general design style (15.2\%). Specifically, for size,
medium-sized robots received the highest utility\footnote{Zero-centred
  utilities---utilities mean-centred within each attribute---are
  reported throughout.} (+0.83), while large robots scored lower (-0.34)
and small robots received the lowest score (-0.48). For location, the
bedroom received the highest utility (+0.29), followed by the living
room (-0.08) and the kitchen (-0.20), suggesting a moderate preference
for robots in personal living spaces. For design style, an
anthropomorphic appearance was rated highest (+0.25), followed by an
industrial design (-0.01) and a functional appearance (-0.24). Regarding
locomotion, wheel-based movement was preferred (+0.33), closely followed
by leg-based movement (+0.25), with track-based movement rated
substantially lower (-0.58). Figure~\ref{fig-results} summarises
attribute importance and part-worth utilities derived from the
conditional logit model.

For the evaluation of the software and the presented stimuli, the
software received high overall ratings across all evaluation criteria.
Participants reported high satisfaction with the survey interface and
navigation (M=4.89, SD=0.32), usefulness of the presented images
(M=4.64, SD=0.56), and the clarity of the presented alternatives
(M=4.75, SD=0.55). While still positive, the lowest mean score was
recorded for the ability to imagine the technology in one's own life
(M=3.98, SD=1.03). This item also showed the highest variance,
suggesting more diverse opinions regarding personal application compared
to the high consensus on the system's technical usability and visual
aesthetics. This may be related to participants' individual associations
with care needs and warrants further investigation beyond the scope of a
proof-of-concept study.

\emph{Conclusion:} The 55 participants performed 440 choice tasks, and
each participant evaluated a subset of over 50 different visual stimuli
generated by a current text-to-image model using a shared base prompt.
The results suggest that participants systematically focus on specific
attributes and levels, as evidenced by the dominant role of robot size
and locomotion, and the comparatively smaller influence of location and
design style on choices.

The export bundle with generated stimuli, design, and anonymised
response data is openly available at \url{https://osf.io/6cqhx/}.

\section{Discussion}\label{discussion}

The integration of generative AI into conjoint (and other) studies spans
several functions. Some AI-assisted functions concern research design
decisions: specifically, the suggestion of attributes and levels. These
suggestions are produced by a language model queried with the survey
title and description, and they may be plausible and well-formed.
However, they must not be adopted without critical review and should
rather be considered as a proof of concept, brainstorming support, or to
quickly demonstrate the conjoint methodology to students. A model
trained on general text will produce attribute suggestions that are
topically plausible but may miss domain-specific constructs, omit
theoretically important dimensions, conflate correlated attributes, or
choose levels that lack discriminant validity. For real research
projects, the selection of attributes in a conjoint study is a
theoretical act. It commits the researcher to a claim that these
dimensions are the relevant axes of variation for the phenomenon under
study, that they are mutually independent in the sense required by the
additive utility model, and that the selected levels adequately span the
relevant range. These commitments cannot be delegated to a language
model.

Other AI-assisted functions concern stimuli production: the generation
of scenario texts and images from pre-specified profiles. Here the
epistemic situation is different. The researcher has already made the
theoretically grounded decisions: the attributes, levels, design
structure, and framing instructions are all determined before generation
begins. The language or image model's task is to instantiate a
structured profile as a natural-looking stimulus, analogous to the work
of a research assistant who translates a stimulus specification into
readable prose or a visual designer who produces mock-ups from a brief.
The base prompt is the researcher's instrument; the generated outputs
are realisations of that instrument across the design space. Provided
the base prompt is carefully constructed and the generated stimuli are
reviewed for quality, this use of generative AI does not compromise the
methodological integrity of the study. It substantially reduces the cost
and time required to produce textual and visual stimuli at the scale
demanded by conjoint designs, and it enables stimuli formats that would
otherwise be prohibitively expensive to create.

However, the prompts used to generate stimuli, including the base prompt
and any level annotations, are therefore part of the study's methodology
and should be reported and archived alongside the generated stimuli, as
supported by GenerativeConjoint's export format.

\subsection{Limitations and Future
Work}\label{limitations-and-future-work}

The current implementation has limitations.

First, the AI-generated stimuli are produced by a single call per
profile without participant-level variation; all participants see the
same generated text or image for a given profile. This means that any
idiosyncratic properties of a generated stimulus (unusual phrasing,
visual artefacts) affect all respondents equally, which is a systematic
rather than random source of measurement error. While the tool enables
checking and regenerating individual stimuli, future work could explore
generating multiple stimuli per profile and randomly assigning them to
participants, allowing the stimulus as a random effect to be modelled
out.

Second, the platform serves as a proof of concept for integrating
generative AI into scientific workflows in the social sciences and
focussed on text and image generation. Future additions may include
additional stimuli types such as generated music or videos. The latter
may support stronger evocation and engagement with the presented
scenarios, though it is unclear how well participants can compare videos
presented side-by-side.

Third, the quality of generated stimuli depends on the base prompt and
annotations, but more importantly on the model used (especially for
image generation). The platform provides AI-assisted prompt optimisation
and feedback functions, and individual stimuli can be inspected and
regenerated from the Inspect Design page. However, there is currently no
automated quality check for generated outputs; assessing whether a
stimulus faithfully and consistently represents its intended profile
remains a manual task for the researcher.

Fourth, the platform currently supports only the standard forced-choice
conjoint format. As conjoint is a family of methods, extensions to
best-worst scaling, ranking-based conjoint, or the inclusion of a
no-choice option would broaden its applicability.

Fifth, analysis support is currently limited to the part-worth-utilities
view and thus requires researchers to create their own analysis scripts
for deeper analysis. Integration of more analyses in the interface may
reduce the additional tooling required after data collection. Yet, the
currently lightweight interface may suffer from adding the variety of
different research lenses that ranges from the plain analysis of
part-worth utilities to latent class analyses and the derivation of
participant profiles.

Lastly, this study on care robot preferences must be viewed primarily as
a proof-of-concept for generating and using integrated visual stimuli in
conjoint-based decision tasks and the estimates should be treated as
exploratory. It is not intended as a definitive investigation into
public perceptions of robots within Ambient Assisted Living (AAL)
contexts. While informed by domain expertise, the study lacks the
theoretical grounding, statistical power\footnote{Sufficient for
  studying attribute preferences but too limited for deeper analyses,
  creating participant profiles, or studying interactions.}, and control
variables and validation necessary to support meaningful conclusions or
specific policy and design recommendations.

\section{Conclusion}\label{conclusion}

This paper presented GenerativeConjoint, an open-source conjoint survey
platform that extends conventional tabular conjoint methodology with
generative AI stimuli modes. The platform covers the full research
lifecycle from design generation through participant data collection to
research-data-management-compliant export, and it does so without
requiring commercial licences or per-response fees. By integrating
generative AI it simplifies survey creation and, more importantly, makes
integrated textual and visual stimuli derived from abstract profiles
practically accessible; it broadens the methodological repertoire
available to researchers who wish to study preferences in richer
representational contexts.

The integration of generative AI in the platform reflects a considered
distinction between functions where AI assistance is epistemically safe
and those where it is not. AI-assisted generation of stimuli from
researcher-specified profiles and base prompts is a legitimate and
valuable tool for simplifying scientific workflows. AI-assisted
suggestion of attributes and levels is a brainstorming or teaching aid
that must be subjected to rigorous expert review. Keeping this
distinction visible to researchers---either through interface design
choices or explicit documentation---is as important as the technical
capability itself.

The platform is available at
\url{https://github.com/braunerphilipp/GenerativeConjoint/} as
open-source. The export bundle with generated stimuli, and anonymised
response data from the example study is openly available at
\url{https://osf.io/6cqhx/}.

\section*{Acknowledgements}\label{acknowledgements}
\addcontentsline{toc}{section}{Acknowledgements}

The author thanks all participants of the proof-of-concept study for
their time and contributions. The thoughtful and critical feedback from
Julia Offermann, Eva-Maria Schomakers, Lena Oden, Felix Glawe, and Ralf
Philipsen on earlier drafts of this work substantially sharpened both
the argument and the presentation of this work. The author also thanks
Martina Ziefle for her curiosity and inspiring ideas that ignited this
research.

\section*{AI Use Declaration}\label{ai-use-declaration}
\addcontentsline{toc}{section}{AI Use Declaration}

Generative AI tools were used in several stages of this work. An AI
coding assistant by Claude Code supported the development of the
GenerativeConjoint platform. The platform's generative AI
pipeline---large language models for textual stimuli and a text-to-image
model for visual stimuli---produced the stimuli used in the
proof-of-concept study; the LLM-facing level annotations that
parameterise these models were drafted with AI assistance (OpenAI) and
validated by the author (the research model with its attributes and
levels was \emph{not} generated by AI). AI assistance was further used
in drafting and editing this manuscript, particularly the technical
details were documented by AI based on the code. The author reviewed and
validated all AI-assisted outputs, confirms that the research design,
analytical framework, and interpretations are the author's own, and
takes full responsibility for the content of this manuscript.

\section*{References}\label{references}
\addcontentsline{toc}{section}{References}

\protect\phantomsection\label{refs}


\begin{CSLReferences}{0}{0}
\bibitem[\citeproctext]{ref-chaudhry2021crackdowns}
\CSLLeftMargin{{[}1{]} }%
\CSLRightInline{S. Chaudhry, M. Dotson, and A. Heiss, {``Who cares about
crackdowns? {Exploring} the role of trust in individual philanthropy,''}
\emph{Global Policy}, vol. 12, no. S5, pp. 45--58, 2021, doi:
\href{https://doi.org/10.1111/1758-5899.12984}{10.1111/1758-5899.12984}.}

\bibitem[\citeproctext]{ref-ryan2000conjoint}
\CSLLeftMargin{{[}2{]} }%
\CSLRightInline{M. Ryan and S. Farrar, {``Using conjoint analysis to
elicit preferences for health care,''} \emph{BMJ}, vol. 320, no. 7248,
pp. 1530--1533, 2000, doi:
\href{https://doi.org/10.1136/bmj.320.7248.1530}{10.1136/bmj.320.7248.1530}.}

\bibitem[\citeproctext]{ref-brauner2025mapping}
\CSLLeftMargin{{[}3{]} }%
\CSLRightInline{P. Brauner, F. Glawe, G. L. Liehner, L. Vervier, and M.
Ziefle, {``Mapping public perception of artificial intelligence:
{Expectations}, risk--benefit tradeoffs, and value as determinants for
societal acceptance,''} \emph{Technological Forecasting and Social
Change}, vol. 220, p. 124304, 2025, doi:
\href{https://doi.org/10.1016/j.techfore.2025.124304}{10.1016/j.techfore.2025.124304}.}

\bibitem[\citeproctext]{ref-offermann2020co2}
\CSLLeftMargin{{[}4{]} }%
\CSLRightInline{J. Offermann-van Heek, K. Arning, A. Sternberg, A.
Bardow, and M. Ziefle, {``Assessing public acceptance of the life cycle
of {CO}\(_2\)-based fuels: {Does} information make the difference?''}
\emph{Energy Policy}, vol. 143, p. 111586, 2020, doi:
\href{https://doi.org/10.1016/j.enpol.2020.111586}{10.1016/j.enpol.2020.111586}.}

\bibitem[\citeproctext]{ref-green1978conjoint}
\CSLLeftMargin{{[}5{]} }%
\CSLRightInline{P. E. Green and V. Srinivasan, {``Conjoint analysis in
consumer research: Issues and outlook,''} \emph{Journal of Consumer
Research}, vol. 5, no. 2, pp. 103--123, 1978, doi:
\href{https://doi.org/10.1086/208721}{10.1086/208721}.}

\bibitem[\citeproctext]{ref-orme2010getting}
\CSLLeftMargin{{[}6{]} }%
\CSLRightInline{B. K. Orme, \emph{Getting started with conjoint
analysis: Strategies for product design and pricing research}, 2nd ed.
Research Publishers, 2010.}

\bibitem[\citeproctext]{ref-sawtooth2024lighthouse}
\CSLLeftMargin{{[}7{]} }%
\CSLRightInline{Sawtooth Software, {``Lighthouse studio.''} Sawtooth
Software, Inc., Provo, UT, 2024. Available:
\url{https://www.sawtoothsoftware.com/lighthouse-studio}}

\bibitem[\citeproctext]{ref-traets2020generating}
\CSLLeftMargin{{[}8{]} }%
\CSLRightInline{F. Traets, D. G. Sanchez Arteta, and M. Vandebroek,
{``Generating optimal designs for discrete choice experiments in {R}:
The idefix package,''} \emph{Journal of Statistical Software}, vol. 96,
no. 3, pp. 1--41, 2020, doi:
\href{https://doi.org/10.18637/jss.v096.i03}{10.18637/jss.v096.i03}.}

\bibitem[\citeproctext]{ref-aizaki2012basic}
\CSLLeftMargin{{[}9{]} }%
\CSLRightInline{H. Aizaki, {``Basic functions for supporting an
implementation of choice experiments in {R},''} \emph{Journal of
Statistical Software}, vol. 50, no. 2, pp. 1--24, 2012, doi:
\href{https://doi.org/10.18637/jss.v050.c02}{10.18637/jss.v050.c02}.}

\bibitem[\citeproctext]{ref-louviere1982design}
\CSLLeftMargin{{[}10{]} }%
\CSLRightInline{J. J. Louviere and D. A. Hensher, {``On the design and
analysis of simulated choice or allocation experiments in travel choice
modelling,''} \emph{Transportation Research Record}, vol. 890, pp.
11--17, 1982.}

\bibitem[\citeproctext]{ref-train2009discrete}
\CSLLeftMargin{{[}11{]} }%
\CSLRightInline{K. E. Train, \emph{Discrete choice methods with
simulation}, 2nd ed. Cambridge University Press, 2009. doi:
\href{https://doi.org/10.1017/CBO9780511805271}{10.1017/CBO9780511805271}.}

\bibitem[\citeproctext]{ref-kuhfeld2010marketing}
\CSLLeftMargin{{[}12{]} }%
\CSLRightInline{W. F. Kuhfeld, \emph{Marketing research methods in
{SAS}}. SAS Institute, 2010.}

\bibitem[\citeproctext]{ref-green1996individualized}
\CSLLeftMargin{{[}13{]} }%
\CSLRightInline{P. E. Green and A. M. Krieger, {``Individualized hybrid
models for conjoint analysis,''} \emph{Management Science}, vol. 42, no.
6, pp. 850--867, 1996, doi:
\href{https://doi.org/10.1287/mnsc.42.6.850}{10.1287/mnsc.42.6.850}.}

\bibitem[\citeproctext]{ref-ryan2008using}
\CSLLeftMargin{{[}14{]} }%
\CSLRightInline{M. Ryan, K. Gerard, and M. Amaya-Amaya, Eds.,
\emph{Using discrete choice experiments to value health and health
care}. Springer, 2008. doi:
\href{https://doi.org/10.1007/978-1-4020-5753-3}{10.1007/978-1-4020-5753-3}.}

\bibitem[\citeproctext]{ref-bridges2011conjoint}
\CSLLeftMargin{{[}15{]} }%
\CSLRightInline{J. F. P. Bridges \emph{et al.}, {``Conjoint analysis
applications in health --- a checklist: A report of the {ISPOR} {Good
Research Practices for Conjoint Analysis Task Force},''} \emph{Value in
Health}, vol. 14, no. 4, pp. 403--413, 2011, doi:
\href{https://doi.org/10.1016/j.jval.2010.11.013}{10.1016/j.jval.2010.11.013}.}

\bibitem[\citeproctext]{ref-hainmueller2014causal}
\CSLLeftMargin{{[}16{]} }%
\CSLRightInline{J. Hainmueller, D. J. Hopkins, and T. Yamamoto,
{``Causal inference in conjoint analysis: Understanding multidimensional
choices via stated preference experiments,''} \emph{Political Analysis},
vol. 22, no. 1, pp. 1--30, 2014, doi:
\href{https://doi.org/10.1093/pan/mpt024}{10.1093/pan/mpt024}.}

\bibitem[\citeproctext]{ref-hainmueller2015hidden}
\CSLLeftMargin{{[}17{]} }%
\CSLRightInline{J. Hainmueller and D. J. Hopkins, {``The hidden
{American} immigration consensus: A conjoint analysis of attitudes
toward immigrants,''} \emph{American Journal of Political Science}, vol.
59, no. 3, pp. 529--548, 2015, doi:
\href{https://doi.org/10.1111/ajps.12138}{10.1111/ajps.12138}.}

\bibitem[\citeproctext]{ref-schomakers2021smart}
\CSLLeftMargin{{[}18{]} }%
\CSLRightInline{E.-M. Schomakers, H. Biermann, and M. Ziefle, {``Users'
preferences for smart home automation---{I}nvestigating aspects of
privacy and trust,''} \emph{Telematics and Informatics}, vol. 64, p.
101689, 2021, doi:
\href{https://doi.org/10.1016/j.tele.2021.101689}{10.1016/j.tele.2021.101689}.}

\bibitem[\citeproctext]{ref-Ladak2024}
\CSLLeftMargin{{[}19{]} }%
\CSLRightInline{A. Ladak, J. Harris, and J. R. Anthis, {``Which
artificial intelligences do people care about most? A conjoint
experiment on moral consideration,''} in \emph{Proceedings of the CHI
conference on human factors in computing systems}, in CHI '24. ACM,
2024, pp. 1--11. doi:
\href{https://doi.org/10.1145/3613904.3642403}{10.1145/3613904.3642403}.}

\bibitem[\citeproctext]{ref-kluber2022appearance}
\CSLLeftMargin{{[}20{]} }%
\CSLRightInline{K. Klüber and L. Onnasch, {``Appearance is not
everything---{P}referred feature combinations for care robots,''}
\emph{Computers in Human Behavior}, vol. 128, p. 107128, 2022, doi:
\href{https://doi.org/10.1016/j.chb.2021.107128}{10.1016/j.chb.2021.107128}.}

\bibitem[\citeproctext]{ref-keil2024watching}
\CSLLeftMargin{{[}21{]} }%
\CSLRightInline{M. Keil, L. Vervier, P. Brauner, and M. Ziefle, {``Will
you be watching me? {A} conjoint-based study on employee attitudes
toward personal data usage in smart factories,''} \emph{International
Journal of Human--Computer Interaction}, vol. 41, no. 16, pp.
10024--10044, 2024, doi:
\href{https://doi.org/10.1080/10447318.2024.2430494}{10.1080/10447318.2024.2430494}.}

\bibitem[\citeproctext]{ref-yeon2024conjoint}
\CSLLeftMargin{{[}22{]} }%
\CSLRightInline{J. Yeon, Y. Jung, Y. Baek, D. Lee, J. Shin, and W. Y.
Chung, {``User preferences on a generative {AI} user interface through a
choice experiment,''} \emph{International Journal of Human--Computer
Interaction}, vol. 41, no. 12, pp. 7626--7637, 2024, doi:
\href{https://doi.org/10.1080/10447318.2024.2400379}{10.1080/10447318.2024.2400379}.}

\bibitem[\citeproctext]{ref-vriens1998verbal}
\CSLLeftMargin{{[}23{]} }%
\CSLRightInline{M. Vriens, G. H. Loosschilder, E. Rosbergen, and D. R.
Wittink, {``Verbal versus realistic pictorial representations in
conjoint analysis with design attributes,''} \emph{Journal of Product
Innovation Management}, vol. 15, no. 5, pp. 455--467, 1998, doi:
\href{https://doi.org/10.1111/1540-5885.1550455}{10.1111/1540-5885.1550455}.}

\bibitem[\citeproctext]{ref-scarpa2009modelling}
\CSLLeftMargin{{[}24{]} }%
\CSLRightInline{R. Scarpa, T. J. Gilbride, D. Campbell, and D. A.
Hensher, {``Modelling attribute non-attendance in choice experiments for
rural landscape valuation,''} \emph{European Review of Agricultural
Economics}, vol. 36, no. 2, pp. 151--174, 2009, doi:
\href{https://doi.org/10.1093/erae/jbp012}{10.1093/erae/jbp012}.}

\bibitem[\citeproctext]{ref-sylcott2016effect}
\CSLLeftMargin{{[}25{]} }%
\CSLRightInline{B. Sylcott, S. Orsborn, and J. Cagan, {``The effect of
product representation in visual conjoint analysis,''} \emph{Journal of
Mechanical Design}, vol. 138, no. 10, 2016, doi:
\href{https://doi.org/10.1115/1.4034085}{10.1115/1.4034085}.}

\bibitem[\citeproctext]{ref-auspurg2014factorial}
\CSLLeftMargin{{[}26{]} }%
\CSLRightInline{K. Auspurg and T. Hinz, \emph{Factorial survey
experiments}, vol. 175. in Quantitative applications in the social
sciences, vol. 175. SAGE, 2015. doi:
\href{https://doi.org/10.4135/9781483398075}{10.4135/9781483398075}.}

\bibitem[\citeproctext]{ref-zwerina1996general}
\CSLLeftMargin{{[}27{]} }%
\CSLRightInline{K. Zwerina, J. Huber, and W. F. Kuhfeld, {``A general
method for constructing efficient choice designs,''} SAS Institute,
1996.}

\bibitem[\citeproctext]{ref-Duyne2002}
\CSLLeftMargin{{[}28{]} }%
\CSLRightInline{D. K. V. Duyne, J. Landay, and J. I. Hong, \emph{The
design of sites: Patterns, principles, and processes for crafting a
customer-centered web experience}. USA: Addison-Wesley Longman
Publishing Co., Inc., 2002.}

\bibitem[\citeproctext]{ref-abdi2018scoping}
\CSLLeftMargin{{[}29{]} }%
\CSLRightInline{J. Abdi, A. Al-Hindawi, T. Ng, and M. P. Vizcaychipi,
{``Scoping review on the use of socially assistive robot technology in
elderly care,''} \emph{BMJ Open}, vol. 8, no. 2, p. e018815, 2018, doi:
\href{https://doi.org/10.1136/bmjopen-2017-018815}{10.1136/bmjopen-2017-018815}.}

\bibitem[\citeproctext]{ref-bradwell2021design}
\CSLLeftMargin{{[}30{]} }%
\CSLRightInline{H. L. Bradwell, G. E. Aguiar Noury, K. J. Edwards, R.
Winnington, S. Thill, and R. B. Jones, {``Design recommendations for
socially assistive robots for health and social care based on a large
scale analysis of stakeholder positions,''} \emph{Health Policy and
Technology}, vol. 10, no. 3, p. 100544, 2021, doi:
\href{https://doi.org/10.1016/j.hlpt.2021.100544}{10.1016/j.hlpt.2021.100544}.}

\end{CSLReferences}
\end{document}